 \documentclass[10pt,prl,twocolumn]{revtex4}

 \usepackage{amssymb}
\usepackage{graphicx}
\usepackage{lipsum}
\usepackage{caption}
\usepackage{amsmath}
\usepackage{bbold}
\usepackage{bm}

\newcommand{\HH}{\mathcal{H}}

\begin{document}
       \title{Quantum optical lattices for emergent many-body phases of ultracold atoms}
         \author{Santiago F.  Caballero-Benitez}
        \email[]{santiago.caballerobenitez@physics.ox.ac.uk}
        \author{Igor B. Mekhov}   \affiliation{ 
 University of Oxford, Department of Physics, Clarendon Laboratory, Parks Road, Oxford OX1 3PU, UK}

\begin{abstract}
Confining ultracold gases in cavities creates a paradigm of quantum trapping potentials. We show that this allows to bridge models with global collective and short-range interactions as novel quantum phases possess properties of both. Some phases appear solely due to quantum light-matter correlations. Due to global, but spatially structured, interaction, the competition between quantum matter and light waves leads to multimode structures even in single-mode cavities, including delocalized dimers of matter-field coherences (bonds), beyond density orders as supersolids and density waves. 
\end{abstract}

       \maketitle

Ultracold atoms trapped in optical lattices (OLs) enable to study quantum many-body phases with undeniable precision and target problems from several disciplines~\cite{Lewenstein}. Such optical potentials can be complicated, but are prescribed, i.e., they are created by external lasers and are not sensitive to atomic phases. This limits the range of obtainable states. Self-consistent light-matter states can be obtained, when scattered light modifies the trapping potential itself. This was achieved by trapping a Bose-Einstein condensate (BEC) inside an optical cavity~\cite{EsslingerNat2010,HemmerichScience2012,ZimmermannPRL2014}, which dramatically enhances the light-matter coupling, thus making the influence of reemission light comparable to that of external lasers. Such ``dynamical potentials"~\cite{RitschRMP} enabled the structural Dicke phase transition and a state with supersolid properties~\cite{EsslingerNat2010}. A key effect observed so far is the dynamical dependence of light intensity (potential depth) on the atomic density. Although, the light becomes dynamical, its quantum properties are still not totally exploited as works on atomic motion in quantum light were limited to few atoms~\cite{EPJD08,VukicsNJP2007,KramerRitschPRA2014,RitschPRA2015}. Effects in dynamical potentials are analogous to semiclassical optics, where atomic excitations are quantum, while light is still classical. As the light and BEC are quantum objects, the quantum fluctuations of both were studied~\cite{DomokosPRA2014,EsslingerStrFact2015}, however, the fundamental reason of the structural phase transition can be traced back to the dynamical self-organization predicted~\cite{Ritsch2002} and observed~\cite{Vuletic2003} with thermal atoms and classical light. For single-mode cavities, dynamical light-matter coupling was shown to lead to several effects~\cite{Larson,Morigi,Hofstetter,Reza,Morigi2} yet to be observed with bosons and with non-interacting fermions~\cite{Keeling,Strack2,Chen}.  Recently an optical lattice inside a cavity has been realized~\cite{2015Esslinger}. Multimode cavities extend the range of quantum phases further~\cite{LevNaturePhys,Strack,Muller2012, KramerRitschPRA2014, Kollar2015}. 
\begin{figure}[htbp!]
\begin{center}
\includegraphics[width=0.48\textwidth]{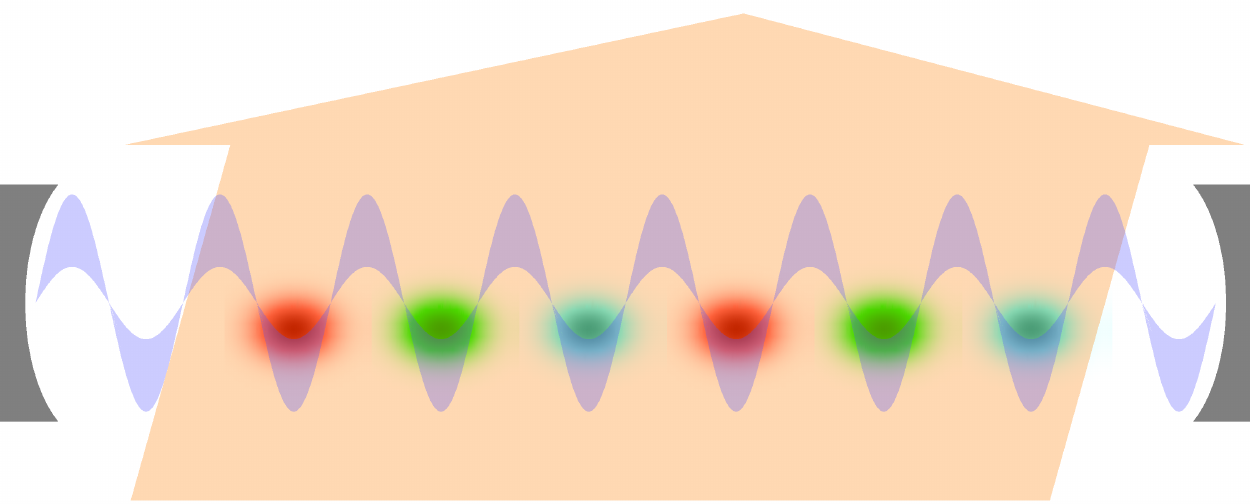}
\end{center}
\captionsetup{width=0.48\textwidth,justification=centerlast,font=small}
\caption{{Cold atoms trapped in an optical lattice subject to a quantum potential created by the light inside a single-mode cavity.} The unsharp potential contour schematically depicts quantum fluctuations of light, which induce the light-matter correlations. The cavity can be a standing- or traveling-wave. Different colors represent atoms corresponding to different light-induced spatially structured atomic modes.}
\label{FQP}
\end{figure}

We show that, even in a single-mode cavity, the quantum potential~\cite{MekhovNP2007,EPJD08,MekhovRev} leads to significant many-body effects beyond semiclassical ones. Multimode spatial patterns of matter fields arise due to symmetry breaking resulting from the competition between imposed global light structure and standard local processes (tunneling and on-site interactions). We demonstrate that the efficient competition is  achieved due to the ability to structure the global interaction at a microscopic scale consistent with the lattice period. Such a competition in turn leads to novel many-body states, which are not limited to density-induced orders as in previous studies, but also represent long-range patterns of matter-field coherences (bonds~\cite{Lauchli2014}), leading, e.g., to far delocalized dimers, trimers, etc.  Importantly, we prove that our approach bridges models with global collective and short-range interactions, as new quantum phases possess properties of both, going beyond Dicke, Lipkin-Meshkov-Glick \cite{ParkinsPRL2008} and other simple spin-1/2 models. Recently, nontrivial spatial patterns were obtained with classical atoms and light~\cite{LabeyrieNaturePhotonics2014}. Our work will assist to extend such efforts in the interdisciplinary field of optomechanics towards quantum multimode systems~\cite{RMP2014Optomech}. The mechanisms we suggest, provide a general framework and a new set of tools, inaccessible in setups using classical OLs. This will strongly expand applications in quantum simulations. It will allow exploring fundamental issues concerning emergence of multimode generalizations of strongly correlated phases, such as gapped superfluids~\cite{Wen} and density waves~\cite{Gruner} as well as their interplay, giving rise to quantum solids~\cite{Pupillo}. The light-induced structure is similar to multi-component nonlinear sigma models ubiquitous in analog models of high-energy~\cite{Zohar, Zoller}, condensed matter~\cite{Auerbach,Bloch}, and relativistic~\cite{Witten} physics. Dimer phases can be used as building blocks for quantum spin-liquids simulations~\cite{Balents}.

We consider atoms trapped in an OL inside single-mode cavity with the mode frequency $\omega_c$ and decay rate $\kappa$ in off-resonant scattering (see ~Fig. \ref{FQP}). The pump light with the amplitude  $\Omega_p$ (in units of the Rabi frequency) and frequency $\omega_p$ ($\Delta_c=\omega_p-\omega_c$) illuminates atoms in a plane transverse to the cavity axis, but not necessarily at $90^\circ$.   The atoms couple with cavity mode via the effective coupling strength $g_2= g \Omega_p/(2\Delta_a)$, where  $g$ is the light-matter coupling coefficient and $\Delta_a$ is the detuning between the light and atomic resonance~\cite{MekhovRev,Wojciech,Gabriel}. This can be described by the Hamiltonian $\HH=\HH^b+\HH^{a}+\HH^{ab}$, where $\HH^b$ is the regular Bose-Hubbard (BH) Hamiltonian~\cite{Fisher, Dieter,supp}. The light is described by  $\HH^{a}=\hbar\omega_c\hat a^\dagger \hat a$ and the light-atom interaction is~\cite{MekhovRev}:
\begin{equation}
\HH^{ab}=g_2^*\hat a\hat F^\dagger+g_2\hat a^\dagger\hat F
\end{equation}
with
$
\hat F= \hat D+\hat B$. $\hat D=\sum_{j}J_{j,j}\hat n_j$ is the diagonal coupling of light to on-site densities, $\hat B=\sum_{\langle i,j\rangle}J_{i,j}( \hat b^\dagger_i\hat b^{\phantom{\dagger}}_j+h.c.)$ is the off-diagonal coupling to the inter-site densities reflecting matter-field interference, or bonds~\cite{Wojciech}. The sums go over illuminated sites $N_s$, $\langle i,j\rangle$ denotes nearest neighbour pairs. The operators $\hat a^\dagger$  ($\hat a$) create (annihilate) photons in the cavity, while $b_i^\dagger$ ($\hat b_i$) correspond to bosonic atoms at site $i$. $\HH^{ab}$  is a consequence of the quantum potential seen by atoms on top of BH model given by a classical OL with the hopping amplitude $t_0$ and on-site interaction $U$. 

The spatial structure of light gives a natural basis to define the atomic modes, as the coupling coefficients $J_{i,j}$ can periodically repeat in space and are calculated from the Wannier functions~\cite{Walters}~see~\cite{supp}. The symmetries broken in the system are inherited from such a periodicity: all atoms equally coupled to light belong to the same mode, while the ones coupled differently belong to different modes. We define operators corresponding to modes $\varphi$: $\hat F=\sum_\varphi\hat D_\varphi+\sum_{\varphi'}\hat B_{\varphi'}$, where
\begin{eqnarray}
\hat D_\varphi&=&J_{D,\varphi}\hat N_\varphi,\;\textrm{with}\;  \hat N_\varphi=\sum_{i\in\varphi}\hat n_i,
\\
\hat B_{\varphi'}&=&J_{B,\varphi'} \hat S_{\varphi'},\;\textrm{with}\; \hat S_{\varphi'}=\sum_{\langle i,j\rangle\in\varphi'}( \hat b^\dagger_i\hat b^{\phantom{\dagger}}_j+h.c.).
\end{eqnarray}
Thus, we replaced the representation of atomic operator $\hat F$ as a sum of microscopic on-site and inter-site contributions by the smaller sum of macroscopically occupied global modes with number density, $\hat N_\varphi$, and bond, $\hat S_\varphi$, operators.The structures of density and bond modes can be nearly independent from each other. To be precise, for the homogeneous scattering in a diffraction maximum, $J_{i,j}=J_B$ and $J_{j,j}=J_D$, one spatial mode is formed. When light is scattered in the main diffraction minimum (at $90^{\circ}$ to the cavity axis), the pattern of light-induced modes alternates sign as in the staggered field, $J_{i,j}=J_{j,i}=(-1)^jJ_B$ and $J_{j,j}=(-1)^jJ_D$. This gives two spatial density modes (odd and even sites) and, as we will show, four bond modes. The density and bond modes can be decoupled by choosing angles such that $J_D=0$ (by shifting the probe with respect to classical lattice thus concentrating light between the sites and assuring the zero overlap between Wannier and mode functions) or $J_B=0$~\cite{Wojciech}. Beyond this, additional modes get imprinted by pumping light at different angles such that each $R$-th site or bond scatters light with equal phases and amplitudes. This generates multimode structures of $R$ density modes~\cite{Thomas, Gabriel} and $2R$ bond modes. The prominent example of self-organization~\cite{Ritsch2002,Morigi,Hofstetter,Reza} is a special case of two density modes, while macroscopic effects related to the higher density modes and any bond modes have not been addressed so far.

In general, the light and matter are entangled~\cite{MekhovNP2007,EPJD08,PRA2009,LP09,LP10,LP11}. In the steady state of light, it can be adiabatically eliminated and the full light-matter state can be then reconstructed as we show in~\cite{supp}. The effective atomic Hamiltonian~\cite{EPJD08,MekhovRev,Morigi} is   
\begin{eqnarray}
 \HH^b_\mathrm{eff}=\HH^b+\frac{g_{\mathrm{eff}}}{2}(\hat F^\dagger \hat F+\hat F \hat F^\dagger),
 \label{effmodel}
\end{eqnarray}
where $g_{\mathrm{eff}}=\Delta_c|g_2|^2/ (\Delta_c^2+\kappa^2)$. A key physical processes is that the ground state is reached [i.e. the energy (\ref{effmodel}) is minimized], when the system adapts (self-organize) in such a way that the light scattering term is maximized for $g_{\mathrm{eff}}<0$, and minimized for $g_{\mathrm{eff}}>0$. New terms beyond BH Hamiltonian give the effective long-range light-induced interaction between density and bond modes: 
\begin{eqnarray}
\hat F^\dagger \hat F+\hat F \hat F^\dagger&=&
\sum_{\varphi,\varphi'}
[\gamma_{\varphi,\varphi'}^{D,D}
\hat N_\varphi^{\phantom{*}}
\hat N_{\varphi'}^{\phantom{*}}
+\gamma_{\varphi,\varphi'}^{B,B}
\hat S_\varphi^{\phantom{*}}\hat S_{\varphi'}^{\phantom{*}}
\nonumber
\\
&+&\gamma_{\varphi,\varphi'}^{D,B}(
\hat N_\varphi^{\phantom{*}}\hat S_{\varphi'}^{\phantom{*}}
+
\hat S_{\varphi}^{\phantom{*}}\hat N_{\varphi'}^{\phantom{*}})],
\label{mdecomp}
\end{eqnarray}
where $\gamma^{\nu,\eta}_{\varphi,\varphi'}=(J_{\nu,{\varphi}}^* J^{\phantom{*}}_{\eta,{\varphi'}}+c.c.)$. Thus, any symmetry broken by the light modes imprints the structure on the interaction of atomic modes. 

Fundamentally, Eq. (\ref{mdecomp}) displays the link between global interactions and the interaction resembling typical short-range one (usually appearing between the sites, while here between the modes). Thus, the resulting quantum phase will have properties of both collective and short-range systems. In this language, it is the term  $\hat N_\text{odd}^{\phantom{*}} \hat N_\text{even}^{\phantom{*}}$ that is responsible for supersolid properties of the self-organized state (The standard supersolidity appears due to $\hat n_i^{\phantom{*}} \hat n_{i+1}^{\phantom{*}}$ interaction.) Our general approach enables to go far beyond typical global models (e.g. Dicke and Lipkin-Meshkov-Glick \cite{ParkinsPRL2008}) due to spatial structuring the global interaction thus assuring its effective competition with the short-range ones, {  even in a single mode cavity in contrast to~\cite{LevNaturePhys,Strack,Muller2012, KramerRitschPRA2014}}. The bond interaction can be easily identified as pseudo-spin interaction via the Schwinger mapping \cite{Auerbach}. In addition, some components can be non-Abelian.

We decompose (\ref{mdecomp}) in mean-field contributions and fluctuations:
\begin{eqnarray}
\hat F^\dagger\hat F+\hat F\hat F^\dagger&=&\langle\hat F^\dagger\rangle\hat F^{\phantom{\dagger}}+\langle\hat F\rangle \hat F^\dagger+\delta\hat F^\dagger\hat F.
\end{eqnarray}
The last term $\delta\hat F^\dagger\hat F$ originates from the quantum light-matter correlations, underlying the quantumness of OL. Other terms originate from the dynamical but classical light, when the semiclassical approximation $\hat a\hat F^\dagger=\langle\hat a\rangle \hat F^\dagger$ holds. Decorrelating operators at different sites, we obtain a mean-field theory that has nonlocal coupling between the matter modes and is local in fluctuations. For $\delta\hat D^\dagger\hat D$ these reduce to on-site number fluctuations. Importantly, this corresponds to the purely light-induced effective on-site interaction of atoms beyond the standard BH term. For $\delta\hat B^\dagger\hat B$, light-matter correlations include radically new terms beyond BH model: fluctuations of the order parameter and density coupling between neighboring sites, which appear due to two and four point quantum atomic correlations~\cite{supp}. In contrast to previous works, we will show non-negligible effects due to such terms, putting forward the quantumness of OL.

When the ground state of $\HH^b_\mathrm{eff}$ is achieved by maximizing scattering ($g_\mathrm{eff}<0$), a strong classical light emerges and small fluctuations can be neglected. In principle, even in the strong-light case, the light quantumness can play a role, because self-organized states can be in a superposition of several patterns and different light amplitudes are correlated to them~\cite{EPJD08}. Nevertheless, in a realistic case with dissipation, the system quickly collapses to one of the semiclassical states~\cite{EsslingerPRL11}. We will show that quantum fluctuations play a key role in the opposite case, where scattering is minimized ($g_\mathrm{eff}>0$).  Here no classical light builds up and light fluctuations design the emergence of novel phases. To underline key phenomena, we will consider cases with either density or bond modes. 

Scattering at $90^\circ$, one breaks the translational symmetry.
Hence, the system can support density waves (DWs) and novel bond orders. The simultaneous occurrence of SF and DW orders is a supersolid (SS) phase~\cite{Pupillo}. SS  and DW have been predicted due to classical maximized scattering~\cite{Hofstetter,Reza}. { We show that  for weak pump, DWs and SSs with only small density imbalance appear at half-integer filling, together with usual MI and SF~\cite{supp,2015Esslinger}. In contrast, above the threshold $|g_{\mathrm{eff}}|N_s>U/2$, we find that DWs and SSs with maximal imbalance are favored, while usual MI and SF are completely suppressed~\cite{supp,2015Esslinger}.} 

For minimized scattering ($g_{\mathrm{eff}}>0$), the classical light cannot build up at all, and quantum fluctuations take the leading role [Fig.\ref{DminJD}(a)].  At fixed density per site, the quantum light-matter fluctuations effectively renormalize the on-site interaction from $U$ to $U+ 2 g_{\mathrm{eff}}J_D^2$ in each component.  Thus, changing the light-matter coupling, one can shift the SF-MI transition point. { This occurs because the light-induced atomic fluctuations now enter the effective Hamiltonian, and these fluctuations need to be suppressed to minimize the energy. This favors MI state for $U$ smaller than that without cavity light, extending MI regions [Fig.\ref{DminJD}(a)]. For incommensurate fillings, SF survives but with smaller (suppressed) fluctuations as well: For convenience, we also plot the boundaries where the superfluid ground state is composed of mainly the two lowest occupation Fock states depending on the filling factor (i.e. components with higher occupations are negligible). Note that with cavity light, the state becomes gapped with respect to adding more than one excitation, thus minimizing fluctuations. 
However there is no phase transition to this peculiar superfluid state~\cite{supp}. }
  Moreover, in a quantum OL, atoms can potentially enter MI even without atomic interaction. This provides absolute control on DW order formation.  In analogous fermionic systems DWs are relevant for the stability of superconducting phases~\cite{HiTc}. 

\begin{figure}[tp!]
\begin{center}
\includegraphics[width=0.48\textwidth]{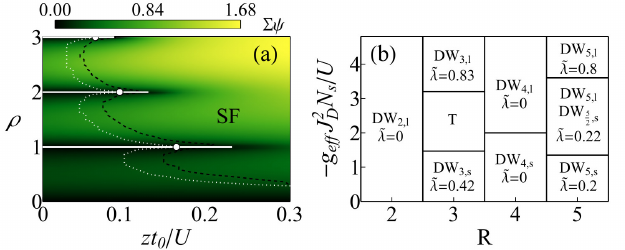}
\end{center}
\captionsetup{width=0.48\textwidth,justification=centerlast,font=small}
\caption{Light scattering from on-site densities. 
(a) Minimized scattering highlighting the quantumness of OL. Total order parameter $\Sigma\psi=|\psi_+|^2+|\psi_-|^2$, where $\psi_\pm$ are the SF order parameters of two modes, white lines correspond to MIs;  dashed lines are boundaries of SF with only two non-negligible Fock components. The system is homogeneous, $\rho_+=\rho_-$ and $\psi_+=\psi_-$. White points are the MI-SF  transition points without cavity light, the SF-MI can be significantly shifted by the quantum OL. 
(b)
Strongly interacting phase diagram for multiple number of modes $R$ (created for different pump angles) at half-filling. Quantum phases have a period of density-density correlations $q$ (in units of
lattice period), $\mathrm{DW}_{q,l}$ ($\mathrm{DW}_{q,s}$), $l$($s$) denotes large (small) density imbalance DW, $\tilde \lambda$ is the SF fraction~\cite{Mahan,Yang,supp}. DW and SF order depend on the effective light-matter interaction strength $g_\mathrm{eff}$ and $R$. Horizontal lines denote phase boundaries between quantum phases.   
 Parameters: (a)  scattering at  $90^\circ$, $g_\mathrm{eff}= 10U/N_s$, the boundaries are for $g_\mathrm{eff}= 10U/N_s$ and $0$; $J_D=1.0$, $J_B=0$, $N_s=100$, $z=6$. (b) scattering at various angles defining the mode number $R$, $J_D=1.0$, $J_B=0$, $t_0=0$, $N_s=100$.}
\label{DminJD}
\end{figure}

{  Scattering at angles different from $90^\circ$ creates more than $R=2$ atomic modes ~\cite{Thomas,Gabriel}, and the light-imposed coefficients are $J_{j,j}=J_D \chi(j)$, where for traveling waves $\chi(j)=\exp\left({i 2\pi j}/{R} \right)$. Now, multiple terms $\hat N_\varphi^{\phantom{*}}
\hat N_{\varphi'}^{\phantom{*}}$ in Eq. (\ref{mdecomp}) become important. We present a phase diagram for various scattering angles (inducing  $R$ modes) for strong on-site interaction with maximal light scattering [Fig.\ref{DminJD}(b)]. The $R$-mode induced pattern competes with on-site interaction modifying the density distribution. Therefore, multiple DWs of period  $R$ can coexist with SF, forming multicomponent SSs. 
Surprisingly, at half-filling for $R > 2$ odd, SS exists.  As $|g_\mathrm{eff}|$ changes from zero, checkerboard insulators form for even $R$ while different kinds of SS exist for odd $R$. The on-site interaction limits atomic fluctuations producing gapped SF components when SS exists. As $R$~increases, additional DWs with different periods and amplitudes emerge. These form unstable mixed state configurations ($T$) for $R=3$ and stable multicomponent SS for $R=5$. These occur in between regular SS phases with DWs of period $R$ and different amplitude. As the light induces atomic fluctuations, these generate competition between small and large amplitude DWs until saturation in the SF component occurs. The system reaches a configuration, similar to the maximal imbalance DW state described above for scattering at $90^\circ$.
}

\begin{figure}[tp!]
    \centering
    \includegraphics[width=0.48\textwidth]{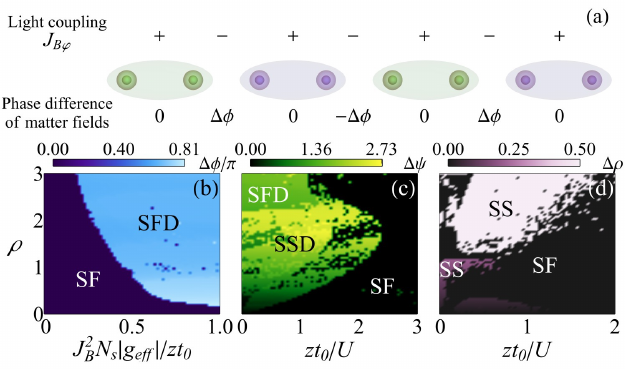}
    \captionsetup{width=0.48\textwidth,justification=centerlast,font=small}
    \caption{{Emergent dimer phases and quantum-light-induced supersolids due to scattering from bonds.} (a) Dimer structure for maximized scattering: matter-field coherence compensates for imposed light-field patterns. (b) Phase diagram for the phase difference $\Delta\phi$ between dimers when light scattering is maximized without on-site interaction. (c) Phase diagram for the difference in order parameters $\psi_{A/B}$ of the dimers $\Delta\psi=||\psi_A|^2-|\psi_B|^2|/2$ when light scattering is maximized with on-site interaction.      (d) Density wave order parameter $\Delta\rho=|\rho_A-\rho_B|/2$  for minimized scattering. The density components $\rho_{A/B}$ correspond to atomic populations of effective light-induced modes. 
     Regions with $\Delta\rho\neq0$ correspond to supersolid phases;  $\Delta\rho=0$ corresponds to superfluid. 
    Parameters:  (b) $U=0$, (c) $g_\mathrm{eff}= -25U/N_s$, (d) $g_\mathrm{eff}=25U/N_s$;   $J_D=0$, $J_B=0.1$, $N_s=100$, and $z=6$.}
    \label{DminJB}
\end{figure}

We find a novel phase transition, when light scattering from the bonds at $90^\circ$ ($J_{j,j+1}=(-1)^jJ_B$, $J_D=0$) is maximized ($g_\mathrm{eff}<0$). Even in the absence of on-site interaction, a transition from normal SF to the superfluid dimer (SFD) state appears. SFD is a SF state in which the complex order parameter has alternating (zero and non-zero) phase difference between pairs of sites, and its amplitude is modulated as well. This occurs because of the competition between the kinetic energy BH terms, which promote a homogeneous SF, with the light-induced interaction that favors SF components with alternating phases across every other site [Fig.\ref{DminJB}(a)]. { The phase of light interference flips from bond to bond [such that the phase difference at neighboring bonds is $\pi$, Fig.\ref{DminJB}(a)]. In the Hamiltonian, this corresponds to alternating signs in front of matter-field coherences between the neighboring sites (i.e. products of complex order parameters, in mean-field treatment). To minimize the energy (and maximize the light scattering) the quantum matter fields self-organize such that the matter-field phase difference between neighboring sites flips as well to compensate for the imposed light-field phase flip. The dimer configurations have high degeneracy as the full many-body ground-state is composed of several equivalent arrangements of the phase pattern in space~\cite{supp}.} The phase diagram is shown in~Fig.\ref{DminJB}(b). Moreover, in the presence of on-site interaction the system supports a transition to the supersolid dimer (SSD) state with modulated densities [Fig.\ref{DminJB}(c)]. {  The on-site interaction  suppresses atomic fluctuations, while as light scattering gets optimized, the density is unable to lock in a homogeneous density pattern leading to additional density imbalance. Therefore, the phase modulation and density modulations coexist simultaneously while atoms retain mobility preventing the stabilization of an insulating phase. } 
Note, that multimode bond structures can have very nontrivial spatial overlap and dimers (and their multimode generalizations as trimers, tetramers, etc., which can be obtained for tilted pump angle) extend over many sites demonstrating the interplay of global and short-range properties. Dimer states can be used as fundamental units to engineer Hamiltonians of quantum spin-liquid states~\cite{Balents}. As $\hat B^\dagger\hat B$ processes enter directly in the effective Hamiltonian with couplings depending on the light geometry and pump amplitude, this makes feasible to achieve analogous physics relaxing the constraint on very low temperatures and large on-site interactions based on second order expansion effective Hamiltonians~\cite{Lewenstein}.
 
Importantly, we prove that there is a SS to SF transition that is solely driven by quantum correlations for minimized light scattering  ($g_{\mathrm{eff}}>0$),Fig.\ref{DminJB}(d). This occurs because the terms due to light-matter correlations in $\hat B^\dagger\hat B$ are not shadowed by semiclassical effects, as there is no classical light build up. Two-point tunneling correlations introduce new terms in the BH model~\cite{supp}, which couple densities at neighboring sites only: $\sum_{\langle i,j\rangle}\hat n_i\hat n_j$,  producing a DW instability even without strong light. {  Density imbalance is energetically favoured and the atoms condense in a nearest-neighbour density pattern, while additional terms in $\HH^b_{\mathrm{eff}}$ favor atomic quantum fluctuations competing with the on-site interaction. }Thus, short range processes induced by the quantumness of light induce the transition.  Such a direct density coupling corresponds stronger to the typical supersolidity scenario, which is under active search~\cite{Pupillo,FerlainoRydberg}.

In experiments with homogeneous BECs both $\kappa$ and the long-range interaction rate are of the same order (either MHz~\cite{PNASEsslinger} or kHz~\cite{PNASHemmerich}), while the depletion rate~\cite{Nagy2010} of the ground state can be made smaller by choosing the detuning such that $|\kappa/\Delta_c|<1$. In our lattice case, the effective light-matter coupling coefficient $g$ rescales as $gJ_{D,B}$, and similarly to the homogeneous systems the evolution can be faster than depletion.

In conclusion, we have shown that quantum optical lattices offer a new tool to engineer nonlocal many-body interactions with light-induced structures. These interactions can break symmetries by design and imprint a pattern that governs the origin of many-body phases. This effectively bridges physics of long-range and short-range interactions. The light and matter are entangled, forming non-trivial light-matter correlated states. We suggested how to generate not only multimode density patterns, but nonlocal patterns of matter-filed coherences as well (in particular, delocalized superfluid and supersolid dimers). Some of the states appear solely due to quantum fluctuations of light and matter, where no classical light can build up.  A pathway to realize our proposal is to combine several recent experimental breakthroughs: a BEC was trapped in a cavity, but without a lattice~\cite{EsslingerNat2010,HemmerichScience2012,ZimmermannPRL2014}, and detection of light scattered from ultracold atoms in OL was performed, but without a cavity~\cite{Weitenberg2011,KetterlePRL2011}. Very recently an optical lattice in a cavity became a reality~\cite{2015Esslinger}. Based on off-resonant scattering and thus being non-sensitive to a detailed atomic level structure, our approach can be extended to other arrays of natural or artificial quantum objects: spins, fermions, molecules (including biological ones)~\cite{LPhys2013}, ions~\cite{Ions2012} , atoms in multiple cavities~\cite{ArrayPolaritons2006}, semiconductor~\cite{SQubits2007}  or superconducting qubits~\cite{JCQubits2009}.  

\begin{acknowledgments}
The work was supported by the EPSRC (EP/I004394/1). 
\end{acknowledgments}

\section*{Supplemental Material}

\textbf{Classical optical lattice model.} The Bose-Hubbard Hamiltonian \cite{Fisher,Dieter} is  
\begin{eqnarray}
\HH^b=-t_0\sum_{\langle i, j\rangle}(\hat b^\dagger_i\hat b^{\phantom{\dagger}}_j+h.c)-\mu\sum_i\hat n_i+\frac{U}{2}\sum_i\hat n_i(\hat n_i-1),
\nonumber
\\
\end{eqnarray}  
where $\langle i,j\rangle$ refers to nearest neighbour pairs,  $b_i^\dagger$ ($\hat b_i$) correspond to creation (annihilation) operators of bosonic atoms at the site $i$ and the atom number operators are $\hat n_i=\hat b^\dagger_i\hat b^{\phantom{\dagger}}_i$. The tunneling amplitude of the bosons is $t_0$, the on-site interaction is $U$ and the chemical potential is $\mu$.  
The effective parameters of the Bose-Hubbard Hamiltonian with the cavity field can be calculated from the Wannier functions and are given by
\begin{eqnarray}
t_0&=&\int w(\mathbf{x}-\mathbf{x}_i)(\nabla^2-V_{OL}(\mathbf{x}))w(\mathbf{x}-\mathbf{x}_j)\mathrm{d}^n x,
\nonumber\\
\\
J_{i,j}&=&\int w(\mathbf{x}-\mathbf{x}_i)u^*_c(x)u_p(\mathbf{x})w(\mathbf{x}-\mathbf{x}_j)\mathrm{d}^n x,
\end{eqnarray}
where $u_{c,p}(\mathbf{x})$ are the cavity and pump mode functions and $w(\mathbf{x})$ are the Wannier functions. The classical optical lattice potential is given by $V_{OL}(\mathbf{x})$. Typical values for the amplitudes of couplings for the standing-wave potential $V_{OL}(x)=V_0 \sin^2(2\pi x/\lambda)$ with $V_0\approx 5E_R$ are: $t_0\approx0.1 E_R$, $|J_B^{i,j}|\approx0.07$,$|J_D^j|\approx0.85$, where $E_R$ is the recoil energy. This has been calculated using real Wannier functions finding the maximally localised generalised Wannier states of the classical optical lattice using the MLGWS code \cite{Walters}.

\textbf{Light-induced interaction decomposition.} 
New terms beyond BH Hamiltonian give the effective long-range light-induced interaction between density and bond modes: 
\begin{eqnarray}
\hat F^\dagger \hat F+\hat F \hat F^\dagger&=&
\sum_{\varphi,\varphi'}(J_{D,{\varphi}}^* J^{\phantom{*}}_{D,{\varphi'}}+c.c.)\hat N_\varphi^{\phantom{*}}
\hat N_{\varphi'}^{\phantom{*}}
\nonumber
\\
&+&
(J_{B,{\varphi}}^* J^{\phantom{*}}_{B,{\varphi'}}+c.c.)\hat S_\varphi^{\phantom{*}}\hat S_{\varphi'}^{\phantom{*}}
\nonumber
\\
&+&(J_{D,{\varphi}}^* J^{\phantom{*}}_{B,{\varphi'}}+c.c)\hat N_\varphi^{\phantom{*}}\hat S_{\varphi'}^{\phantom{*}}
\nonumber
\\
&+&(J_{B,{\varphi}}^* J^{\phantom{*}}_{D,{\varphi'}}+c.c)\hat N_{\varphi}^{\phantom{*}}\hat S_{\varphi'}^{\phantom{*}}.
\label{mdecomp}
\end{eqnarray}

We separate light matter-correlations and dynamical terms in $\hat F^\dagger\hat F$  performing  on-site mean-field. The $\hat D^\dagger\hat D$ term can be expanded as
\begin{eqnarray}
\hat D^\dagger\hat D+\hat D\hat D^\dagger&\approx&\sum_{\varphi,\varphi'}(J_{D,\varphi}^*J_{D,\varphi'}^{\phantom{*}}+c.c.)\langle\hat N_{\varphi'}^{\phantom{*}}\rangle(2\hat N_\varphi-\langle\hat N_\varphi^{\phantom{*}}\rangle),
\nonumber
\\
&+&\delta \hat D^\dagger\hat D
\label{DS1}
\\
\label{DS2}
\delta \hat D^\dagger\hat D&=&2\sum_\varphi |J_{D,\varphi}|^2\delta\hat N_\varphi^2,
\\
\delta\hat N_\varphi^2&=&\sum_{i\in\varphi}(\hat n_i-\rho_i)^2,
\end{eqnarray}
where $\langle\hat N_\varphi\rangle=\sum_{i\in\varphi}\rho_i$ is the mean number of atoms in the mode $\varphi$ and $\rho_i=\langle\hat n_i\rangle$ is the mean atom number at site $i$. The first term in Eq. (\ref{DS1}) is due to the dynamical properties of the light field, these terms exhibit nonlocal coupling between light-induced modes.  The terms in (\ref{DS2}) signifies the light-matter correlations and contain the effect due to quantum fluctuations.

The $\hat B^\dagger\hat B$ terms can be expanded as:
\begin{eqnarray}
\hat B^\dagger\hat B+\hat B\hat B^\dagger&\approx&\sum_{\varphi,\varphi'}(J_{B,\varphi}^*J_{B,\varphi'}^{\phantom{*}}+c.c.)\langle\hat S_{\varphi'}^{\phantom{*}}\rangle(2\hat S_\varphi-\langle\hat S_\varphi^{\phantom{*}}\rangle)
\nonumber
\\
&+&\delta\hat B^\dagger\hat B,
\label{BS1}
\\
\delta\hat B^\dagger\hat B&=&2\sum_\varphi |J_{B,\varphi}|^2\delta \hat S_\varphi^2,
\label{BS2}
\end{eqnarray}
\begin{eqnarray}
\delta \hat S_\varphi^2&=&
\sum_{\langle i,j \rangle\in\varphi}(
(\hat b_i^\dagger\hat b_j^{\phantom{\dagger}}+h.c.-\langle\hat b_i^\dagger\hat b_j^{\phantom{\dagger}}+h.c.\rangle)^2
\nonumber
\\
&+&
\sum_{\langle i,j,k \rangle\in\varphi}
\big(
\hat b^\dagger_i\hat b^\dagger_k(\hat b^{\phantom{\dagger}}_j)^2
+(\hat b^\dagger_j)^2\hat b^{\phantom{\dagger}}_i\hat b^{\phantom{\dagger}}_k
+
2\hat n_i^{\phantom{\dagger}}\hat b^\dagger_k\hat b^{\phantom{\dagger}}_j
+\hat b^\dagger_k\hat b^{\phantom{\dagger}}_j
\nonumber\\
&&\phantom{\sum_{\langle i,j,k}
}-(\hat b^\dagger_k\hat b^{\phantom{\dagger}}_j+\hat b^\dagger_i\hat b^{\phantom{\dagger}}_k+h.c.)\langle\hat b_i^\dagger\hat b_j^{\phantom{\dagger}}+h.c.\rangle\big),
\label{BS2}
\end{eqnarray}
where  $\langle i,j,k \rangle$ refers to $i$,$j$ nearest neighbours and $k$ is a nearest neighbour to the pair $\langle i,j\rangle$. The first term in (\ref{BS1}) is due to the dynamical properties of the light field and (\ref{BS2}) are due to light-matter correlations. These are basically all the possible 4 point correlations and tunneling processes between nearest neighbours, as higher order tunneling processes have much smaller amplitudes. In principle, additional terms might be considered in the above expansion and their generalization is straight forward by removing the restriction over the sums beyond nearest neighbours. The on-mode  light matter correlations $\hat B_\varphi^\dagger \hat D_\varphi^{\phantom{\dagger}}+\hat D_\varphi^\dagger\hat B_\varphi^{\phantom{\dagger}}$ reduce to on-site covariances per mode. In the above we have decorrelated products of operators at different sites such that  
$\hat \xi_i\hat \nu_j\approx\langle\hat \xi_i\rangle\hat \nu_j+\langle\hat \nu_j\rangle\hat \xi_i-\langle\hat \xi_i\rangle\langle\hat \nu_j\rangle$, where $\hat\xi$ or  $\hat\nu$ are  combinations of $\hat b^\dagger$ or $\hat b$ operators at a given site.  The expectation value of the bond operators reduces to $\langle \hat S_{\varphi}\rangle=\sum_{\langle i, j\rangle\in\varphi}(\psi_i^\dagger\psi^{\phantom{\dagger}}_j+\psi_j^\dagger\psi_i^{\phantom{\dagger}})$, which is the sum of products of order parameters at nearest neighbour sites in the light-induced mode $\varphi$. In principle additional incident or output light-modes might be considered by trivially adding a sub-index and sum over additional incident  or output modes, this is equivalent to include additional light induced bands in the effective Hamiltonian. The technique can be fully transferred to multimode cavities and the coupling of additional cavities in the system.
 
\textbf{Effective Hamiltonians.} The effective Hamiltonian considering only diagonal coupling ($\hat D^\dagger\hat D+\hat D\hat D^\dagger$) is
\begin{eqnarray}
\HH^b_{\mathrm{eff}}&=&
-t_0\sum_{\langle i, j\rangle}(\hat b^\dagger_i\hat b^{\phantom{\dagger}}_j+h.c)
\\
&+&\sum_{\varphi}\Big[\sum_{i\in\varphi}\left(\frac{U_\varphi}{2}\hat n_i(\hat n_i-1)-2g_{\mathrm{eff}}|J_{D,\varphi}|^2\rho_i\hat n_i\right)
\nonumber
\\
&-&\mu_\varphi\hat N_\varphi-g_{\mathrm{eff}}c_{D,\varphi}\Big],
\nonumber\\
\mu_\varphi&=&\mu-g_{\mathrm{eff}}\eta_{D,\varphi},
\\
U_\varphi&=&U+2g_{\mathrm{eff}}|J_{D,\varphi}|^2,
\end{eqnarray}
with $\eta_{D,\varphi}=\sum_{\varphi'}(J_{D,\varphi}^*J^{\phantom{*}}_{D,\varphi'}+c.c.)\langle \hat N_{\varphi'}\rangle-|J_{D,\varphi}|^2$ and $c_{D,\varphi}=\sum_{\varphi'}(J_{D,\varphi}^*J^{\phantom{*}}_{D,\varphi'}+c.c.)\langle\hat N_{\varphi'}\rangle \langle \hat N_{\varphi}\rangle/2-|J_{D,\varphi}|^2\sum_{i\in\varphi}\rho_i^2$. The many-body interaction $U_\varphi$ and the chemical potential $\mu_\varphi$ inherit the pattern induced by the quantum potential that depends on light-induced mode structure given by $\varphi$. 

In the case of only off-diagonal bond scattering  ($\hat B^\dagger\hat B+\hat B\hat B^\dagger$) , we have
\begin{eqnarray}
\HH^b_{\mathrm{eff}}&=&\sum_{\varphi}\Big[-t_\varphi\hat S_\varphi+ g_{\mathrm{eff}}|J_{B,\varphi}|^2\delta \hat S_\varphi^2-g_{\mathrm{eff}}c_{B,\varphi}\Big]
\nonumber\\
\\
&+&\sum_i\left(\frac{U}{2}\hat n_i(\hat n_i-1)-\mu\hat n_i\right),
\\
t_\varphi&=&t_0-g_{\mathrm{eff}}\eta_{B,\varphi},
\end{eqnarray}
with 
$\eta_{B,\varphi}=\sum_{\varphi'}(J_{B,\varphi}^*J^{\phantom{*}}_{B,\varphi'}+c.c.)\langle \hat S_{\varphi'}\rangle$ and $c_{B,\varphi}=\sum_{\varphi'}(J_{B,\varphi}^*J^{\phantom{*}}_{B,\varphi'}+c.c.)\langle\hat S_{\varphi'}\rangle \langle \hat S_{\varphi}\rangle/2$. The effective tunneling amplitude $t_\varphi$ couples the SF components of all the light-induced modes $\varphi$. The terms due to $\delta \hat S_\varphi^2$ induce nontrivial coupling between nearest neighbour sites and lead to the formation of a density wave instability with more than one light-induced mode.

The effective coupling parameters $\eta_{D/B,\varphi}$  depend implicitly on all the light induced modes giving nonlocal coupling via the expectation values of the operators $\hat S_{\varphi}$ and $\hat N_{\varphi}$.

\begin{figure}[tp!]
\begin{center}
\includegraphics[width=0.48\textwidth]{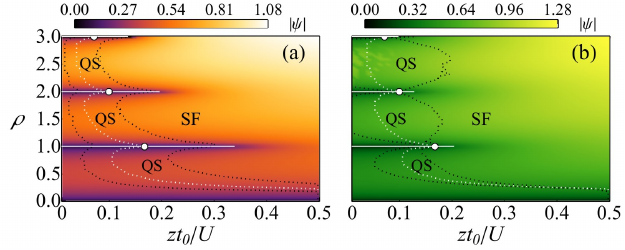}
\end{center}
\captionsetup{width=0.48\textwidth,justification=centerlast,font=small}
\caption{{\bf Modifications of quantum phases due to quantum and semiclassical effects for homogenous scattering}. (a) Phase diagram in terms of SF order parameter $\psi$ at fixed density for density-induced scattering. For minimal (maximal) scattering, MI boundaries (white lines) become extended (shortened) with respect to the transition point without cavity light (white point). This corresponds to suppression (enhancement) of quantum-light-induced atomic fluctuations. The behaviour of boundaries of the gapped QS state is similar (black dotted lines). The white dotted line is for gapless QS without cavity light. (b) Phase diagram for off-diagonal bond-induced scattering. The processes and lines are similar to (a), but arise due to renormalization of the tunnelling amplitude resulting from semiclassical light scattering.
Parameters: (a) $g_\mathrm{eff}=25U/N_s$, the boundaries are for $g_\mathrm{eff}=25U/N_s$, $0$, and $-12.5U/N_s$; $J_D=1.0$, $J_B=0$; (b)  $g_\mathrm{eff}=1.0U/N_s$, the boundaries are for $g_\mathrm{eff}=1.0U/N_s$, $0$, and $-1.0U/N_s$; $J_D=0$, $J_B=0.05$; (a,b) $N_s=100$ and $z=6$ (3D).
\label{Dmax}}
\end{figure}

{\bf Homogeneous scattering at diffraction maximum.}
In the specific case of a single light-induced mode component in the diffraction maxima, we have: $\eta_D=2J_D^2(N_s\rho-1/2)$ and $\eta_B=2zJ_B^2N_s|\psi|^2$ with $\rho=\langle\hat n_i\rangle$ and $\psi=\langle\hat b_i\rangle$ for all sites and $z$ is the coordination number. This gives the effective tunneling amplitude $t_\varphi=t_0-2zg_{\mathrm{eff}}J_B^2N_s|\psi|^2$, the effective chemical potential $\mu_\varphi=\mu-g_{\mathrm{eff}}J_D^2(2(N_s-1)\rho-1)$  [where we have added all onsite density terms], and the effective interaction strength $U_\varphi=U+2g_{\mathrm{eff}}J_{D}^2$. { Density-dependent, but classical light was previously shown to strongly modify the standard phase diagram of Mott insulator (MI) - superfluid (SF) transition \cite{Morigi,Larson}, if plotted via the chemical potential $\mu$. If the phase diagram is plotted via the density, rather than chemical potential, most of such modifications due to the classical light become invisible. Therefore, in this paper we chose to plot the phase diagrams via the density to focus more on the effects produced by the quantum light. For example, at fixed density per site, it is the quantum light-matter correlations that effectively renormalize the on-site interaction from $U$ to $U+ 2 g_{\mathrm{eff}}J_D^2$ [see Fig. \ref{Dmax}(a)]. Thus, changing the light-matter coupling, one can shift the SF-MI transition point due to essentially the quantumness of light.}

Choosing geometry \cite{Wojciech}, one can suppress the density scattering ($J_D=0$) and have all $J_B$'s equal. As no symmetry is broken, the bond self-organization does not emerge, but another semiclassical effect arises: tunneling is enhanced (suppressed) for maximum (minimum) scattering. This modifies the phase diagram (Fig. \ref{Dmax}b), because of nonlinear coupling of the SF order parameter $\psi=\langle\hat b_i\rangle$ to the tunneling amplitude $t_0$, which renormalizes to $t_0- 2 zg_{\mathrm{eff}} J_B^2N_s|\psi|^2$.

{\bf A Peculiar Superfluid: the quantum superposition (QS) state.}
In the presence of cavity light, the gap opens in the SF state because the effective chemical potential depends on the density and renormalizes to $\mu-g_{\mathrm{eff}}J_D^2((2N_s-1)\rho-1)$. Without tunneling for $g_\mathrm{eff}>0$, the energy required to add a particle on top of the ground state is $\Delta E_{\mathrm{QS}}(\rho)=U\rho+g_\mathrm{eff}J_D^2(2N_s\rho+1)$ for incommensurate fillings between MI regions with fillings  $n$ and $n+1$, with  
 \begin{equation}
 \rho=\frac{U }{2g_\mathrm{eff}J_D^2N_s}\left(\frac{\mu}{U} -n\right)
 \end{equation}
  and on-site number fluctuations $\Delta(\hat n_i)=(\rho-n)(1-\rho+n)$. For MI  regions at commensurate density $\rho=n$, the gap is $\Delta E_{\mathrm{MI}}(n)=U n+g_\mathrm{eff}J_D^2(2N_s n+1)$ and $\Delta(\hat n_i)=0$. { This means that only occupations of the lowest particle-hole excitations are allowed in between MI lobes. { Therefore in mean-field approximation this peculiar SF state is made of the quantum superposition of only two lowest Fock components that would satisfy the constraint on the density, instead of a coherent superposition of all possible fillings. Therefore, we use a term quantum superposition (QS) state for such a peculiar SF with a gap.}  The usual MI physics with renormalized  interaction strength occurs at commensurate fillings.} Similarly to SF-MI transition point, the boundaries of QS can be tuned (stabilized) due to the quantum correlations as $U$ is renormalized [Fig.\ref{Dmax}(a)]. The quantum superposition state can be understood as the SF state of minimal atomic fluctuations. { Note that a gap opens in the spectrum of excitations in the system, but because the particle filling is not commensurate with the lattice the system can't form an insulator and the best alternative energetically is the QS. As no symmetry has been broken in the system, the state is smoothly connected to the regular SF state and no phase transition occurs in the usual sense.}

\begin{figure}[tp!]
\begin{center}
\includegraphics[width=0.48\textwidth]{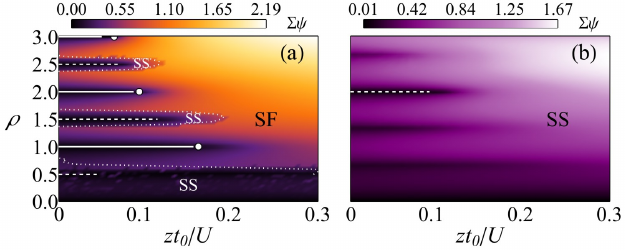}
\end{center}
\captionsetup{width=0.48\textwidth,justification=centerlast,font=small}
\caption{ {\bf Emergence of density wave and supersolid phases in the diffraction minima of light with density coupling  for maximized light scattering from the atoms ($g_{\mathrm{eff}}<0$)}. (a)  Phase diagram for the total order parameter of two density modes $\Sigma\psi=|\psi_+|^2+|\psi_-|^2$, where $\psi_\pm$ are the SF parameters of two modes. The DW order parameter is $\Delta\rho=|\rho_+-\rho_-|/2$ when $\Delta\rho\neq0$ there is DW order ( $\rho_\pm$ are the mean atom numbers per site in each mode). Dashed lines at half-integer fillings correspond to DW insulators ($\Sigma\psi=0$, $\Delta\rho\neq0$), SS state occurs, when $\Sigma\psi\neq 0$ and $\Delta\rho\neq 0$.  White solid lines at integer fillings correspond to MI; dotted lines are the boundaries between SS and SF. (b)  Phase diagram where only giant DW and SS exist above the quantum critical point for light-matter coupling (no SF or MI phases appear). The white points correspond to the classical optical lattice SF to MI transition.The boundaries of MI, SF, and MI can be tuned by the pump-cavity detuning.  Parameters: (a)  $g_\mathrm{eff}=-0.5 U/N_s$, (b) $g_\mathrm{eff}=-1.25U/N_s$, $N_s=100$, $z=6$ (3D).    }
\label{DminJD}
\end{figure}
{\bf Diffraction Minima.}
Scattering light at $90^\circ$, one explicitly breaks the translational symmetry, which gets imprinted on the interaction of modes. In this case, we have  two components in the diffraction minima in diagonal density scattering reduces to  $\eta_{D,\pm}=\pm J_D^2N_s\Delta\rho-J_D^2$ with $\Delta\rho=(\rho_+-\rho_-)/2$, where $\rho_\pm$ correspond to the mean atom numbers of each of the two light-induced modes. Thus, the effective chemical potential is  $\mu_\pm=\mu\pm g_{\mathrm{eff}}J_D^2N_s\Delta\rho-g_{\mathrm{eff}}J_D^2$ and $U_\pm=U+2g_{\mathrm{eff}}J_{D}^2$.  In the absence of bond scattering ($J_{j,j}=(-1)^jJ_D$ $J_{i,j}=0$), the system can support supersolid (SS), density wave (DW) insulators, MI and SF  (Fig. \ref{DminJD}a). The system supports DW insulators present at half-integer fillings surrounded by a SS shell while MI's exist at commensurate filling.  There exists a critical light-matter coupling ($|g_{\mathrm{eff}}|N_s>U/2$), where MI and SF are entirely suppressed, which is a quantum critical point. The system only supports supersolid (SS) and density waves (DW) due to incommensurability  (Fig. \ref{DminJD}b). It forms giant density waves with maximal amplitude constrained by the mean atom number per site and is described by a Devil's staircase \cite{Devils}. Interestingly a DW insulator phase, checkerboard lattice, will appear above the giant DW transition at $\rho=2$, consistent with~\cite{Reza}.

 For off-diagonal coupling, when bond ordering occurs  ($J_{i,j}=(-1)^jJ_B$ $J_{i,i}=0$) the situation is more subtle and this requires four components, 
\begin{eqnarray}
t_{\varphi_q}&=&t_0-g_{\mathrm{eff}}\eta_{B,{\varphi_q}},
\\
\eta_{B,\varphi_q}&=&\frac{z(-1)^{q+1
}J_B^2N_s}{2}\sum_{q'=1}^4(-1)^{q'+1}(\psi_{q'}^*\psi^{\phantom{*}}_{q'+1}+c.c.),
\nonumber\\
\end{eqnarray}
 where the component $q+4$ is the same as $q$. This is the origin of dimer states discussed in the main text.

 \textbf{The full entangled light-matter state.} The full light-matter state can be found using an alternative procedure to \cite{EPJD08, PRA2009} using rotations over the Hilbert space via canonical transformations \cite{Mahan}. We obtain
\begin{equation}
|\Psi\rangle\approx\sum_{\varphi_q}\Gamma^b_{\varphi_q}(t)|\varphi_q\rangle_b|\alpha_{\varphi_q}\rangle_a,
\end{equation}
where the subscript ``$a$" (``$b$") corresponds to the light (matter) part. The coherent state components for the light are:
\begin{equation}
|\alpha_{\varphi_q}\rangle_a=D(\alpha_{\varphi_q})|0\rangle_a
\end{equation}
with $D(\alpha)=\exp(\alpha\hat a^\dagger- \alpha^*\hat a)$, the displacement operator. The coherent state amplitudes are $\alpha_{\varphi_q}=c F_{\varphi_q}$, where $c={g_2}/({\Delta_c+i\kappa})$. These amplitudes depend on the projection $F_{\varphi_q}|\varphi_q\rangle_b=\hat F|\varphi_q\rangle_b$, that corresponds to the  steady state component $|\varphi_q\rangle_b$ of the effective matter Hamiltonian  $ \HH^b_\mathrm{eff}$.  The phase factor is due to the time evolution of the effective matter Hamiltonian: $\hat \Gamma^b(t)=\exp(-i\HH^b_\mathrm{eff} t)$, with projection $\Gamma^b_{\varphi_q}|\varphi_q\rangle_b=\hat\Gamma^b|\varphi_q\rangle_b$.  This is accurate in the limits  $|\kappa/\Delta_c|\ll1$, $|t_0/\Delta_c|\ll1$, $|J_B E_R/\Delta_c|\ll1$. { The ground state of the effective Hamiltonian is $|\Psi\rangle_b=\sum_{q=1}^{f(R)}|\varphi_q\rangle_b$, where the number of components related to the light induced modes is $f(R)$. 

For the multimode DW's ($R\ge 2$) originating from density scattering, the number of components $f(R)\geq 2$ depend on the sub-lattice structure generated. The coherent state amplitudes are $\alpha_{\varphi_q}=cN_{\varphi_q}$ and depend on the projections $N_{\varphi_q}|\varphi_q\rangle_b=\hat D|\varphi_q\rangle_b$.  

Similarly, for dimer states in mean-field approximaton, 
 $\alpha_{\varphi_q}=c B_{\varphi_q}$ with 
\begin{equation} 
B_{\varphi_q}= N_sJ_B[\phi_{q}+\phi_{q+2}-(\phi_{q+1}+\phi_{q+3})\cos(\Delta\phi)]/2,
\end{equation}
while 
 $\phi_{q}=|\psi_{q+1}^*\psi_{q}|$,
 $\Delta\phi=\arg(\psi_{q+2})-\arg(\psi_{q})=\arg(\psi_{q+2})-\arg(\psi_{q+1})$, and $f(R)=4$.
The components $q$ and $q+4$ are the same and $\psi_q=\langle\hat b_q\rangle$ is the order parameter. For SS dimers $\phi_{q}\neq\phi_{q+2}$ and  $\phi_{q+1}\neq\phi_{q+3}$, while for SF dimers $\phi_{q}=\phi_{q+2}$ and $\phi_{q+1}=\phi_{q+3}$.  

We see that both density and bond multimode self-organized states are represented by the superpositions of several macroscopic components. Using the quantum measurement of light, one can project the full states to one or several degenerate components, inducing novel states and dynamics~\cite{Gabriel}. }

\textbf{Numerical methods.} Numerical simulations of the effective Hamiltonian have been carried out by doing spatial mode decoupling theory truncated in Hilbert space using the Gutzwiller ansatz  with self-consistent conditions with light-induced interaction given by Eqs. (\ref{DS1}) and (\ref{BS1}). The problem is solved as a multimode constrained nonlinear optimisation problem in $2R(f+1)$ dimensions and $3R$ constraints with $R$ being the number of light-induced modes and $f$ is the filling factor of the mode occupation. { For our parameter range it is sufficient to consider $f=5$. 
Exact diagonalization calculations are done in the effective Hamiltonian Eq. (4) in the main text in 1D for small lattices in the absence of tunneling. We calculate the largest eigenvalue of the reduced density matrix $\lambda$
scaled by the atom number $N$, $\tilde{\lambda}=\lambda/N > 0$. The condensate fraction estimated by $\tilde\lambda$ is an upper bound to the thermodynamic limit value of the SF fraction in the system~\cite{Mahan,Yang}. Checkerboard insulators have 
$\tilde\lambda=0$ and SS has $\tilde\lambda\neq0$.}

  \end{document}